# Microstructure-controlled vortex phases and two-phase superconductivity in $(TaNb)_{0.7}(HfZrTi)_{0.5}$ revealed by ac magnetostrictive coefficients

Mengju Yuan[1,*], Yuze Xu[2,*], Bin Zhang[1], Jun-Yi Ge[3], Aifeng Wang[1], Mingquan He[1], Yanpeng Qi[2,4,5,a)], Yisheng Chai[1, a)]

[1]*Low Temperature Physics Laboratory, College of Physics & Center of Quantum Materials and Devices, Chongqing University, Chongqing 401331, China*

[2] *State Key Laboratory of Quantum Functional Materials, School of Physical Science and Technology, ShanghaiTech University, Shanghai 201210, China*

[3]*Materials Genome Institute, Shanghai University, Shanghai 200444, China*

[4] *ShanghaiTech Laboratory for Topological Physics, ShanghaiTech University, Shanghai 201210, China*

[5] *Shanghai Key Laboratory of High-resolution Electron Microscopy, ShanghaiTech University, Shanghai 201210, China*

*These authors contributed equally to this work

a)Authors to whom correspondence should be addressed: qiyp@shanghaitech.edu.cn and yschai@cqu.edu.cn

**ABSTRACT.** We investigate flux dynamics in the high-entropy alloy superconductor $(TaNb)_{0.7}(HfZrTi)_{0.5}$ after annealing (as-cast, 500 °C, 550 °C, and 1000 °C) using a sensitive ac composite magnetoelectric method that measures the complex ac magnetostrictive coefficient $(d\lambda/dH)_{ac}$. The resulting vortex phase diagrams show that intermediate annealing (500–550 °C) induces nanoscale clustering, enhances pinning, and produces a pronounced fishtail effect with successive elastic- and plastic-vortex-glass regimes. Flux-jump instabilities appear at 550 °C and persist at 1000 °C, indicating strong pinning and thermomagnetic instability in the low-temperature, low-field regime. Remarkably, the 1000 °C sample exhibits a two-step superconducting response—a double plateau or drop in $d\lambda'/dH$ and two dissipation peaks in $d\lambda''/dH$—demonstrating the coexistence of two superconducting phases with distinct irreversibility and critical-field value. We further show that the resolvability of the two-step $(d\lambda/dH)_{ac}$ signature is governed by the topological connectivity of the phase-separated microstructure, which controls magnetic shielding between the TaNb-rich network and the $(TaNb)_{0.7}(HfZrTi)_{0.5}$ parent phase. These results establish a direct microstructure–vortex-state correlation and provide a route to tailoring flux pinning in chemically complex superconductors via thermal processing.

High-entropy alloys (HEAs), introduced by Yeh and Cantor in 2004[1,2], represent a paradigm shift from traditional alloy design. Unlike conventional alloys based on one or two principal elements, HEAs comprise multiple principal components (typically 4–9, occasionally up to 20) in near-equimolar ratios. Their defining feature is the strong chemical disorder that generates a high configurational entropy. Thermodynamically, the Gibbs free energy of mixing, $\Delta G_{mix} = \Delta H_{mix} - T\Delta S_{mix}$, can be minimized by a large entropy contribution ($\Delta S_{mix}$), stabilizing simple disordered solid-solution phases such as (BCC)[3–8], face-centered cubic (FCC)[9–11], and hexagonal close-packed (HCP)[12,13] structures, where $\Delta H_{mix}$ is the mixing enthalpy. Kinetically, sluggish diffusion suppresses atomic rearrangement during cooling, thereby retaining the disordered solid-solution structure at ambient conditions. These entropy-driven effects collectively give HEAs excellent chemical, physical, and mechanical properties[14,15], making them promise for applications beyond the reach of conventional alloys.

In 2014, superconductivity was first discovered in HEA $Ta_{34}Nb_{33}Hf_8Zr_{14}Ti_{11}$[16], which exhibits type-II superconductivity with a superconducting transition temperature $T_c$ = 7.3 K and an upper critical field $H_{c2}$ ~82 kOe. Subsequently, various HEA superconductors with diverse crystal structures have been reported. Their crystal structures and $T_c$ can often be described by the valence electron concentration (VEC)[17,18], following the empirical Matthias rule[19]. For instance, BCC-type HEA superconductors typically have VEC values between 4.0 and 5.0, reaching a maximum $T_c$ of ~8 K near VEC ≈ 4.7[11,17]. In addition, the HEA concept has also been successfully extended to established superconducting systems, including oxide high-$T_c$ superconductors $REBa_2Cu_3O_{7-x}$ (RE = Y, Dy, Gd, Sm, Eu)[20] and $Nb_3$(Al, Sn, Ge, Ga, Si)[21] compounds, and so on, suggesting a broad applicability of chemical complexity to superconducting materials design.

Among HEA superconductors, the BCC Ta–Nb–Hf–Zr–Ti system [with atomic occupancy schematically shown in Fig. 1(b)] exhibits $T_c$ ≈ 7–10 K. Notably, this system maintains robust zero resistance under extreme hydrostatic pressures up to 190.6 GPa[22], making it both a model platform for high-entropy superconductivity and a promising candidate for extreme-environment applications. At the same time, its superconducting properties are highly microstructure sensitive, with thermal annealing serving as an effective control route. As demonstrated by Vrtnik et al.[23], annealing drives atomic diffusion, relaxing the system toward the minimum $\Delta G_{mix}$. At sufficiently high annealing temperatures $T_a$ (near the melting temperature, $T > 2500\ °C$), the $\Delta S_{mix}$ term dominates, favoring a homogeneous disordered solid solution. In contrast, at intermediate $T_a$ (450–1170 °C), the mixing enthalpy $\Delta H_{mix}$ gains influence, promoting nanostructures like Zr/Hf-rich clusters within a Ta/Nb-rich matrix[24]. This annealing-tuned microstructural evolution is expected to strongly affect flux pinning and the vortex phase diagram, yet systematic studies in this area remain lacking, particularly those that directly track vortex dissipation and phase boundaries in a unified manner.

Recently, Gao et al.[24] systematically studied the microstructural evolution of $(TaNb)_{0.7}(HfZrTi)_{0.5}$ over a wide annealing range (up to 1170 °C), identifying a sequence of regimes with increasing $T_a$: intrinsic chemical disorder (as-cast), a fishtail regime, the onset of flux-jump behavior, and eventual cluster-driven phase separation accompanied by persistent flux jumps [Fig. 1(a)]. Notably, annealing at 450–550 °C induces a pronounced fishtail effect in the magnetization hysteresis loops. Within the Bean model, this effect manifests as a valley followed by a peak in the field ($H$) dependent critical current density $J_c(H)$, defining the characteristic fields $H_{min}$ and $H_{sp}$, respectively. These fields mark transitions between distinct vortex phases: an elastic vortex-glass phase for $H_{min} < H < H_{sp}$, and a plastic vortex-glass phase for $H_{sp} < H < H_{irr}$ ($H_{irr}$: the irreversible field). Analysis of the pinning-force density $F_p$ ($\propto J_c H$) further highlights multiple pinning mechanisms and the complexity of vortex physics in HEA superconductors, likely governed not only by nanoscale precipitation but also by the inherent chemical disorder. However, their vortex phase mapping focused primarily on the as-cast and 500 °C annealed samples. A key unresolved question therefore emerges: as the microstructure evolves systematically with annealing temperature, how does the vortex phase diagram respond and dynamically reconstruct.

To map the complete phase diagram evolution accurately without resorting to complicated transport measurements, this study employs an ac composite magnetoelectric (ME) technique that is highly sensitive to vortex dynamics. By probing the complex ac magnetostrictive coefficient $(d\lambda/dH)_{ac}$ ($\lambda = \Delta L/L$, where $L$ is the sample dimension) under a small ac field excitation, this method enables high-resolution discrimination among vortex states. Specifically, the real and imaginary parts of $(d\lambda/dH)_{ac}$ are sensitive to vortex entry, pinning, and dissipative motion, allowing one to distinguish vortex-solid, vortex-liquid, and non-vortex regimes and to determine key phase boundaries such as the $H_{irr}$ and the upper critical field $H_{c2}$. This approach has been successfully applied to several typical type-II superconductors, such as $YBa_2Cu_3O_{7-x}$, $Nb$[25], and $EuFe_2(As_{1-x}P_x)_2$[26], revealing fine phase structures that are difficult to resolve with conventional measurements. It is therefore particularly suited for analyzing the complex vortex phase diagram and dynamics in HEA superconductors.

Motivated by the phase-region framework reported for $(TaNb)_{0.7}(HfZrTi)_{0.5}$[24], we systematically investigated four representative annealing conditions: as-cast, 500 °C, 550 °C, and 1000 °C. The obtained vortex phase diagrams evolve systematically with $T_a$, reflecting distinct pinning behaviors across regimes. In particular, the $T_a = 500$ °C and 550 °C samples, where clustering develops, exhibit successive elastic and plastic vortex-glass regimes; pronounced flux jumps appear for $T_a = 500$ °C. By contrast, the $T_a = 1000$ °C sample shows clear phase separation together with persistent flux jumps at low temperatures, and it further provides an opportunity to examine how the topology (connectivity/percolation) of the phase-separated microstructure influences the $(d\lambda/dH)_{ac}$ vortex signature.

Polycrystalline sample $(TaNb)_{0.7}(HfZrTi)_{0.5}$ was synthesized by arc melting under an argon atmosphere. To ensure homogeneity, the ingot was remelted and flipped multiple times, with negligible mass loss. The sample was cut into a regular cuboid-like shape, sealed in quartz tubes partially filled with argon, and annealed at specified temperatures ($T_a$ = 500°C, 550°C, 1000°C) for 5 hours. Elemental mapping was performed using an electron backscatter diffraction (EBSD) system equipped with an energy-dispersive X-ray spectroscopy (EDS) detector. Magnetization measurements were carried out in a 7 T Magnetic Property Measurement System (MPMS3, Quantum Design).

For the composite magnetoelectric (ME) measurements, the polycrystalline

(TaNb)$_{0.7}$(HfZrTi)$_{0.5}$ sample was mechanically combined to a [001]-cut single crystal 0.7Pb(Mg$_{1/3}$Nb$_{2/3}$)O$_3$-0.3PbTiO$_3$ (PMN-PT) piezoelectric plate (thickness: 0.2 mm) using a two-component epoxy (EPO-TEK H20E). The PMN–PT plate served as a strain medium, as shown in Fig. 1(c) lower panel. Both the ac excitation field (1 Oe in amplitude) and the dc bias field were applied along the thickness direction of the piezoelectric plate. The ME voltage $V_{ME}$ was measured using a lock-in amplifier (OE1022 DSP) with a commercial sample stick (Multifield Corp.) in a 14 T magnet (Oxford TeslatronPT). $V_{ME}$ is proportional to $(d\lambda/dH)_{ac} = d\lambda'/dH + id\lambda''/dH$ that it has been treated as $(d\lambda/dH)_{ac}$ all over the paper (as shown in Fig. 1(c) lower panel).

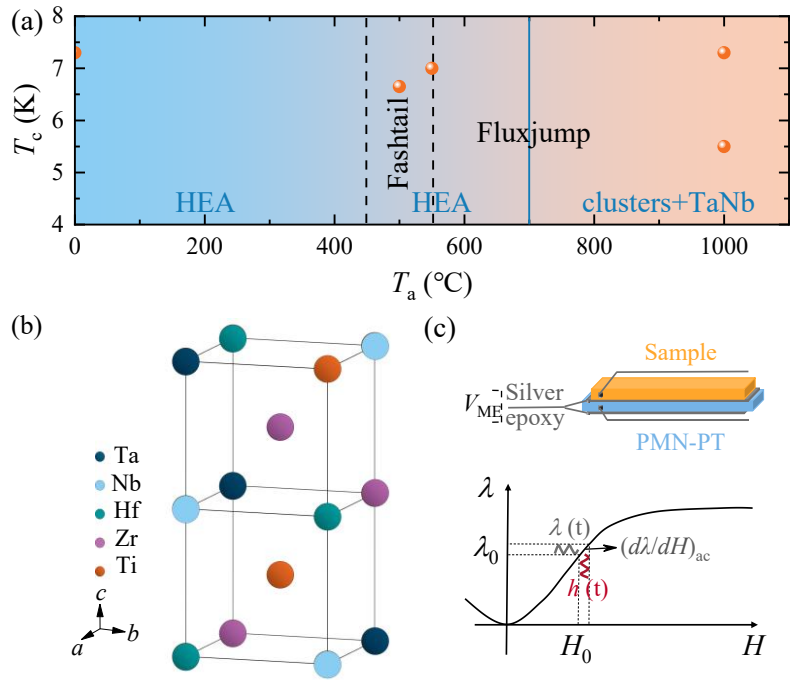

**Fig. 1.** (a) Phase-region summary for samples annealed at different temperatures, adapted from Gao *et al*.[24] The four annealing temperatures studied here are indicated. Orange solid circles mark the superconducting transition temperatures determined from the $(d\lambda/dH)_{ac}$ signal for the samples in this work. (b) Schematic of the BCC crystal structure of the HEA. (c) Schematic of the ac composite magnetoelectric method used for $(d\lambda/dH)_{ac}$ characterization.

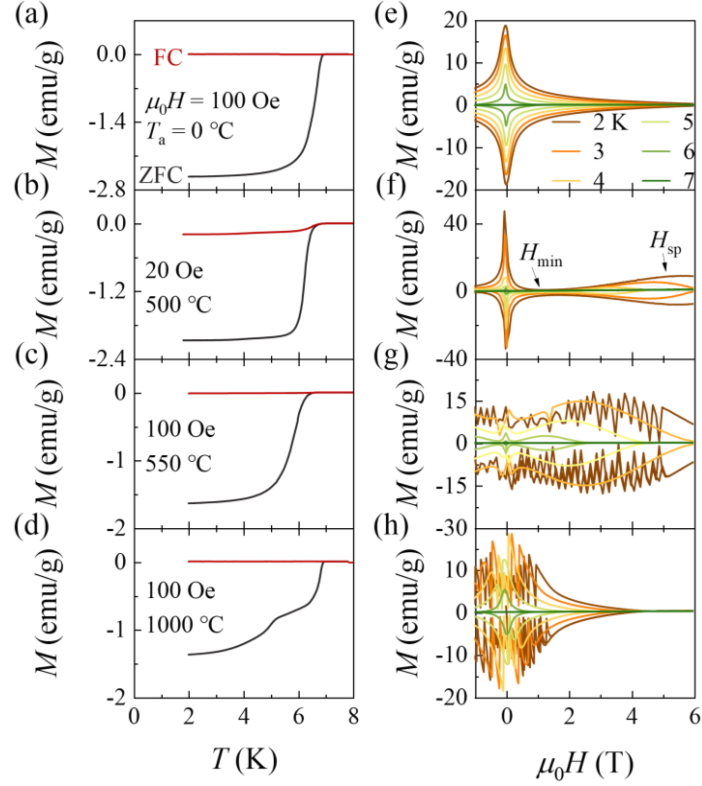

**Fig. 2.** (a-d) Field-cooled and zero-field-cooled magnetization as a function of temperature for samples annealed at different $T_a$, measured in a small dc magnetic field. (e-h) Magnetic hysteresis loops measured at selected temperatures for the corresponding samples, which are associated with the $T_a$ shown on the left. Data for the $T_a = 500$ °C sample are taken from Gao et al[24].

To quickly check the superconductivity of the $(TaNb)_{0.7}(HfZrTi)_{0.5}$ samples annealed at various $T_a$, we first performed systematic magnetization ($M$) measurements on the four samples. Figures 2(a)–(d) show the field-cooled (FC) and zero-field-cooled (ZFC) $M$-$T$ curves measured in a small dc magnetic field. All samples exhibit clear diamagnetic ZFC responses, confirming superconductivity. Notably, the 1000 °C sample shows a two-step transition in ZFC, suggesting the emergence of a secondary superconducting phase. Figures 2(e)–2(h) present isothermal magnetic hysteresis loops ($M$-$H$ loops). All samples show well-centered, symmetric loops typical of bulk pinning, but their detailed behavior depends strongly on $T_a$. The as-cast sample [Fig. 2(e)] exhibits a conventional type-II response. By contrast, the 500 °C sample [Fig. 2(f)] shows a pronounced fishtail effect, with a distinct second peak and characteristic fields

$H_{min}$ = 1.1 T and $H_{sp}$ = 5.15 T at 2 K. With increasing temperature, the fishtail effect weakens, the peak shifts to lower fields, and it vanishes near $T_c$. Similar phenomenology is observed in $Ba_{0.6}K_{0.4}Fe_2As_2$[27] and $YBa_2CuO_{7-\delta}$[28] systems, although the microscopic origin in HEAs can be distinct. For 550 °C [Fig. 2(g)], the fishtail effect coexists with flux jumps: a clear second peak persists below 6 K, while abrupt low-field discontinuities appear at 2–3 K. Such discontinuities are commonly attributed to abrupt flux entry into or expulsion from the superconductor and are often observed in type-II bulk superconductors with strong pinning, high critical current density and under reasonably fast field sweeping rate, including HEAs[29,30]. This coexistence places 550 °C near the boundary between fishtail and flux-jump regimes, consistent with Fig. 1(a). At 1000 °C [Fig. 2(h)], the fishtail effect is absent while flux jumps persist at low temperature, indicating suppression of the pinning mechanism responsible for the second peak under high-temperature annealing. Overall, these magnetization results place the selected samples within the phase-region framework reported previously.

Previous work by Gao et al.[24], reported additional diffraction peaks corresponding to a cubic TaNb phase when $T_a$ exceeds 700 °C. Since TaNb in the cubic phase is also superconducting[31], we performed high-resolution EBSD to confirm the formation and spatial distribution of TaNb in the 1000 °C sample. The EBSD maps reveal a well-defined two-phase mosaic microstructure at the micrometer scale: dark-phase clusters are embedded within a connected bright background [Supplemental Fig. S1(a)]. Elemental mapping [supplementary material Fig. S1(a)] indicates that the dark regions are Hf/Zr-rich, while the bright regions are Ta/Nb-rich, consistent with TaNb-rich precipitates. This supports the assignment of one superconducting transition to the TaNb-rich phase.

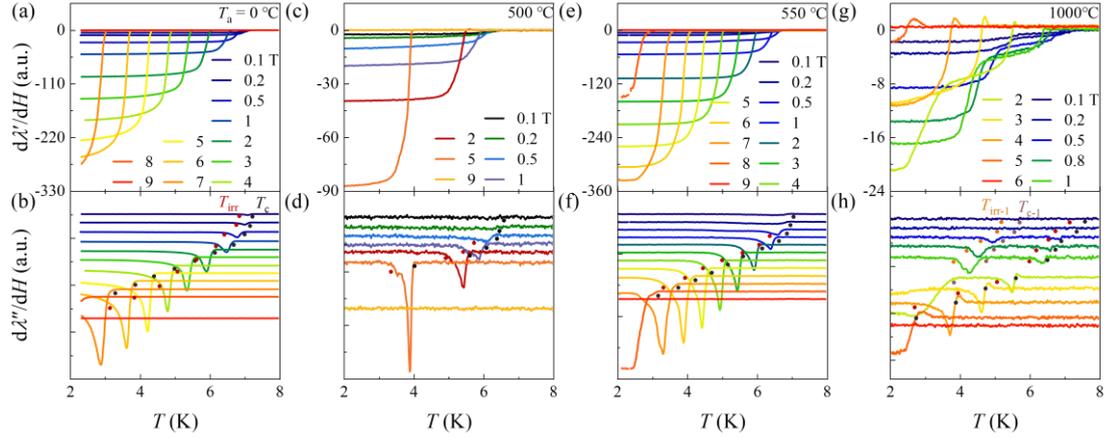

**Fig. 3.** Upper panels (a, c, e, g): Temperature dependence of the real part (d$\lambda'$/d$H$) of (d$\lambda$/d$H$)$_{ac}$ under different dc field-cooled for samples annealed at various $T_a$. Lower panels (b, d, f, h): Corresponding temperature dependence of the imaginary part (d$\lambda''$/d$H$) of the (d$\lambda$/d$H$)$_{ac}$ under the same conditions. Solid circles of different colors represent different transition temperatures. For clarity, all d$\lambda''$/d$H$ curves have been vertically offset.

To systematically investigate vortex dynamics and phase boundaries, we measured (d$\lambda$/d$H$)$_{ac}$ as a function of temperature under selected magnetic fields, as shown in Fig.3. The temperature dependences behaviors of the real (d$\lambda'$/d$H$) and imaginary (d$\lambda''$/d$H$) parts of (d$\lambda$/d$H$)$_{ac}$ for the four samples are roughly consistent with those of a typical type-II superconductor[25]. In the vortex-solid regime, d$\lambda'$/d$H$ generally exhibits a nearly temperature-independent plateau, while d$\lambda''$/d$H$ shows negligible dissipation. In our previous studies of type-II superconductors[25], the low-temperature plateau in d$\lambda'$/d$H$ is closely related to vortex density. Upon warming, a rapid suppression of d$\lambda'$/d$H$ together with a pronounced peak in d$\lambda''$/d$H$ signals entry into a vortex-liquid regime. In the normal state, both d$\lambda'$/d$H$ and d$\lambda''$/d$H$ approach zero. Using these criteria, $T_c$ and $T_{irr}$ are extracted for each sample. Notably, for the 1000 °C sample, d$\lambda'$/d$H$ develops two distinct plateaus upon cooling, while d$\lambda''$/d$H$ shows two dissipation features [Figs. 3(g) and 3(h)]. Together with the two-step ZFC transition and the EBSD results, this directly supports phase separation and attributes the double-plateau/double-dissipation response to coexistence of the (TaNb)$_{0.7}$(HfZrTi)$_{0.5}$ parent phase and a TaNb-rich phase.

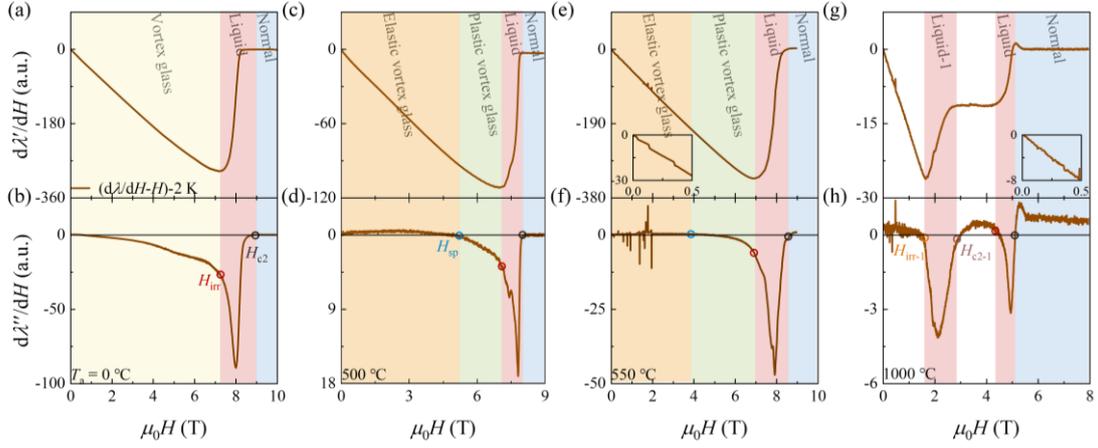

**Fig. 4.** Upper panels (a), (c), (e), (g): Field dependence of the d$\lambda'$/d$H$ at 2 K for samples annealed at various $T_a$. Lower panels (b), (d), (f), (h): Corresponding field dependence of d$\lambda''$/d$H$ under the same conditions. Open circles of different colors represent different magnetic transition fields.

The field-dependent ac magnetostrictive coefficients at selected temperatures were also measured, as shown in Fig. 4 and supplementary material Fig. S2. Systematic evolution of vortex dynamics across different annealing conditions are revealed in both d$\lambda'$/d$H$ and d$\lambda''$/d$H$ at 2 K as a representative (Fig. 4). For the as-cast sample [Fig. 4(a)], the absolute value of d$\lambda'$/d$H$ increases with field as vortex density grows and reaches a maximum at $H_{irr}$. The corresponding d$\lambda''$/d$H$ [Fig. 4(b)] is finite but weak, consistent with a vortex-glass regime. Gao *et al.* reported a vortex glass in the as-cast sample with relatively low $J_c$, implying a weak pinning force $F_p$[24]. Such weak pinning allows slow flux motion slightly below $H_{irr}$, consistent with the gradual field evolution of d$\lambda''$/d$H$. The resulting $H_{irr}$ is consistent with the merging point of the upward and downward sweeping $M$-$H$ curves at elevated temperatures [e.g., Fig. 2(e)], though such correspondence could not be confirmed in the high-field regime at 2 K due to measurement limitations. The consistency across temperatures is further demonstrated in subsequent data (supplementary material Fig. S2). With increasing field, the absolute value of d$\lambda'$/d$H$ decreases and vanishes at the upper critical field $H_{c2}$, where d$\lambda''$/d$H$ also approaches zero. Between $H_{irr}$ and $H_{c2}$, d$\lambda''$/d$H$ displays a pronounced dissipation peak, identifying the vortex-liquid regime.

The 500 °C and 550 °C samples follow the overall field evolution of the as-cast sample but show distinct low-field behavior. Flux-jump features and the characteristic fields $H_{min}$ and $H_{sp}$ extracted from magnetization are clearly reflected in $(dλ/dH)_{ac}$. Flux jumps appear as step-like features in $dλ'/dH$ [inset of Fig. 4(e)] and spike-like features in $dλ''/dH$ [Fig. 4(f)], similar to observations in Nb[25]. In addition, $H_{min}$ corresponds to a clear nonlinear deviation in $dλ'/dH$ (supplementary material Fig. S3), which we attribute to a change in the elastic response of the vortex lattice as it reorganizes into a correlated elastic-glass regime. Above $H_{sp}$, a weak but finite $dλ''/dH$ appears, indicating plastic motion and rearrangement under the driven ac field. As the field increases toward $H_{c2}$, $dλ''/dH$ develops a strong maximum, marking crossover into the vortex-liquid regime.

For the 1000 °C sample, the low-field response resembles that of the 550 °C sample, including flux jumps [inset of Figs. 4(g) and 4(h)]. Below $H_{irr-1}$, $dλ'/dH$ increases approximately linearly with field while $dλ''/dH$ remains negligible, consistent with a vortex-lattice regime of a conventional type-II superconductor[25]. Above $H_{irr-1}$, the absolute value of $dλ'/dH$ shows a distinct two-step drop, and $dλ''/dH$ exhibits two dissipation peaks, providing additional evidence for two coexisting superconducting phases: a TaNb-rich precipitate phase with the smaller irreversible value and the $(TaNb)_{0.7}(HfZrTi)_{0.5}$ parent phase with the larger one. These features persist at elevated temperatures and show no clear sweep-direction hysteresis (supplementary material Fig. S2). Notably, such clear two-phase superconducting transitions are not readily resolved in the conventional $M$-$H$ hysteresis loops in Fig. 2(h).

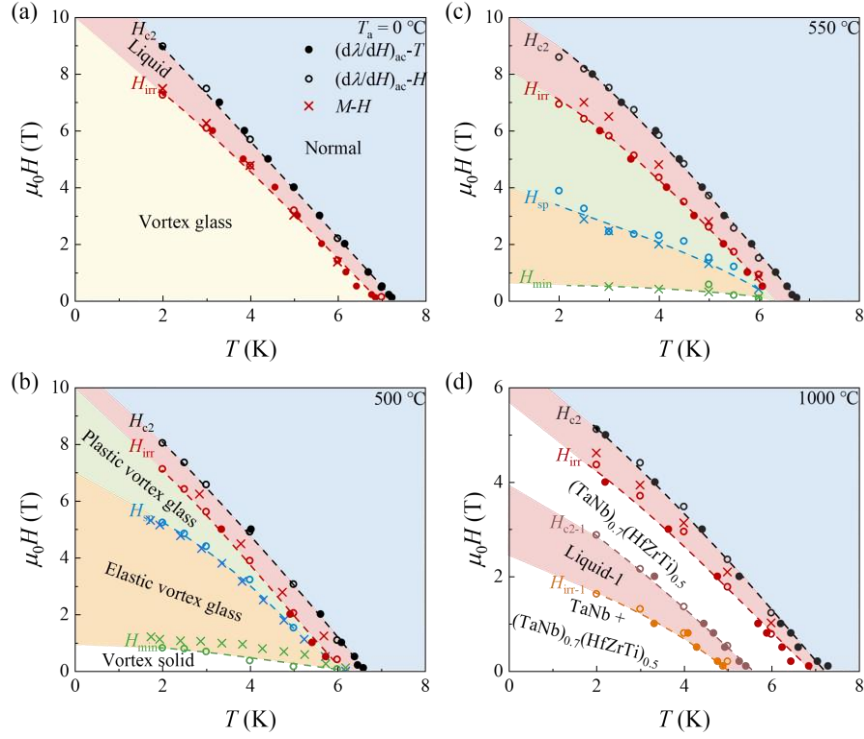

**Fig. 5.** (a)–(d) Magnetic field-temperature ($H$-$T$) vortex phase diagrams of $(TaNb)_{0.7}(HfZrTi)_{0.5}$ annealed at 0 °C, 500 °C, 550 °C, and 1000 °C, respectively. Different colored regions represent distinct phases. The characteristic transition fields and temperatures are determined from: solid circles-$(d\lambda/dH)_{ac}$-$T$ curves, open circles - $(d\lambda/dH)_{ac}$-$H$ curves, and crosses-$M$-$H$ loops.

Based on systematic analysis of the $(d\lambda/dH)_{ac}$ data, we construct the vortex $H$-$T$ phase diagrams for all annealing conditions (Fig. 5). Transition fields extracted from $M$-$H$ loops are also included and are broadly consistent with the $(d\lambda/dH)_{ac}$-derived boundaries. For the as-cast sample [Fig. 5(a)], three regimes are identified: the normal state, separated from the vortex-liquid regime by $H_{c2}$ line, and the vortex liquid, separated from the vortex-glass regime by the $H_{irr}$ line. For the 500 °C and 550 °C samples, clustering-enhanced pinning and the fishtail effect are accompanied by two additional glass regimes: an elastic vortex glass between $H_{min}$ and $H_{sp}$, and a plastic vortex glass between $H_{sp}$ and $H_{irr}$ [Figs. 5(b) and 5(c)]. For the 1000 °C sample, EBSD, $M$-$T$ and $(d\lambda/dH)_{ac}$ measurements consistently indicate two superconducting phases [Fig. 5(d)]. Importantly, the topological connectivity of the two-phase microstructure governs whether both phases are resolvable in $(d\lambda/dH)_{ac}$: if the TaNb-rich phase forms a strongly percolative network, magnetic shielding can suppress the contribution from the parent phase; in contrast, the observed mosaic topology (with parent-phase regions not fully screened) allows the two-step plateaus/peaks to remain visible, enabling simultaneous extraction of two sets of irreversible and critical-field value[32,33].

In summary, controlled annealing in $(TaNb)_{0.7}(HfZrTi)_{0.5}$ provides a direct route to tune microstructure and reconstruct the vortex phase diagram. $(d\lambda/dH)_{ac}$ measurements reveal that intermediate annealing (500–550 °C) induces clustering-enhanced pinning and successive elastic and plastic vortex-glass regimes, with prominent flux jumps at 550–1000 °C. At 1000 °C, annealing drives macroscopic phase separation into $(TaNb)_{0.7}(HfZrTi)_{0.5}$ and TaNb -rich phases with distinct $T_c$ and $H_{irr}$ scales, and the visibility of the resulting two-step $(d\lambda/dH)_a$ signature is governed by the connectivity of the phase-separated microstructure.

See the supplementary material for additional experimental details and data supporting this work.

**ACKNOWLEDGMENTS**

This work was supported by the National Natural Science Foundation of China (Grant Nos. 11674347, 11974065, 51725104, 11774399, 11474330, 52101221, U21A201910), Fundamental Research Funds for the Central Universities (Project No. 2024IAIS-ZX002), the National Key Research and Development Program of China (Grants No. 2018YFA0704200 and No. 2023YFA1406100). Y. S. Chai would like to thank the support from Beijing National Laboratory for Condensed Matter Physics. We would like to thank Ms. Y. Liu at Analytical and Testing Center of Chongqing University for their assistance.

**AUTHOR DECLARATION**

**Conflict of Interest**

The authors have no conflicts to disclose.

## Author Contributions

Mengju Yuan and Yuze Xu contributed equally to this work.

**Mengju Yuan**: Data curation (equal); Methodology (equal); Writing – original draft (equal); Writing – review & editing (equal). **Yuze Xu**: crystal growth (equal); Data curation (equal); Methodology (equal). **Bing Zhang**: Methodology (equal). **Jun-Yi Ge**: Conceptualization (equal); Supervision (equal). **Aifeng Wang**: Conceptualization (equal); Supervision (equal). **Mingquan He**: Conceptualization (equal); Supervision (equal). **Yanpeng Qi**: Conceptualization (equal); Supervision (equal); Writing – review & editing (equal). **Yisheng Chai**: Conceptualization (equal); Supervision (equal); Writing – original draft (equal); Writing – review & editing (equal).

## DATA AVAILABILITY

The data that support the findings of this study are available from the corresponding author upon reasonable request.

# Supplementary Materials

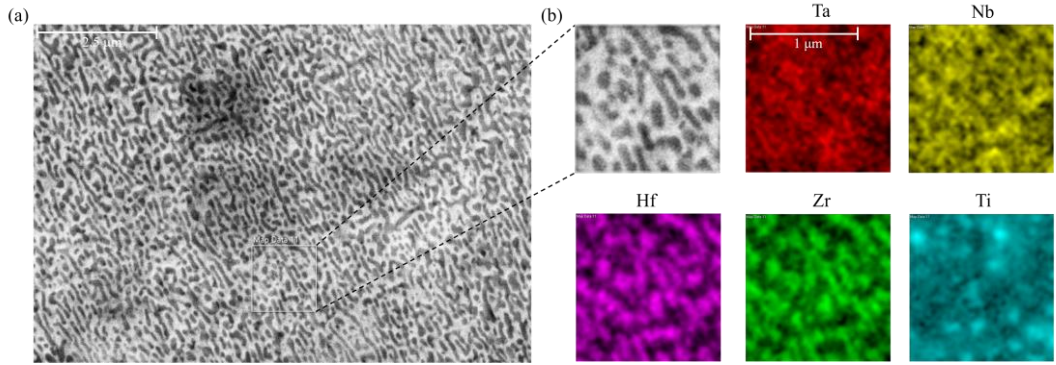

**FIG. S1** (a) Electron image of the overall EBSD-scanned area in the sample with $T_a$ = 1000 °C. (b) Corresponding EBSD orientation map and elemental distribution maps for Ta, Nb, Hf, Zr, and Ti.

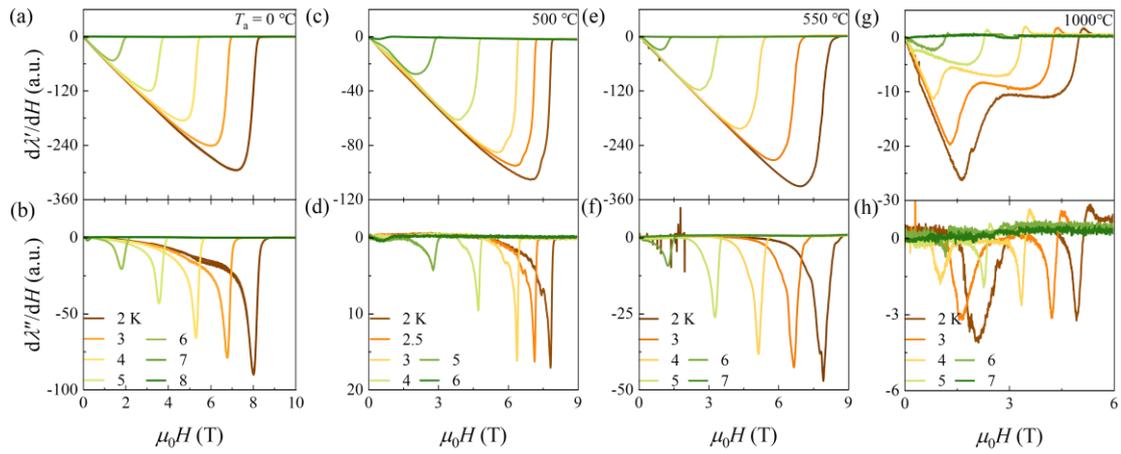

**FIG. S2** Upper panels (a, c, e, g): Field dependence of the $d\lambda'/dH$ at different temperature for samples annealed at various $T_a$. Lower panels (b, d, f, h): Corresponding field dependence of the $d\lambda''/dH$ under corresponding conditions.

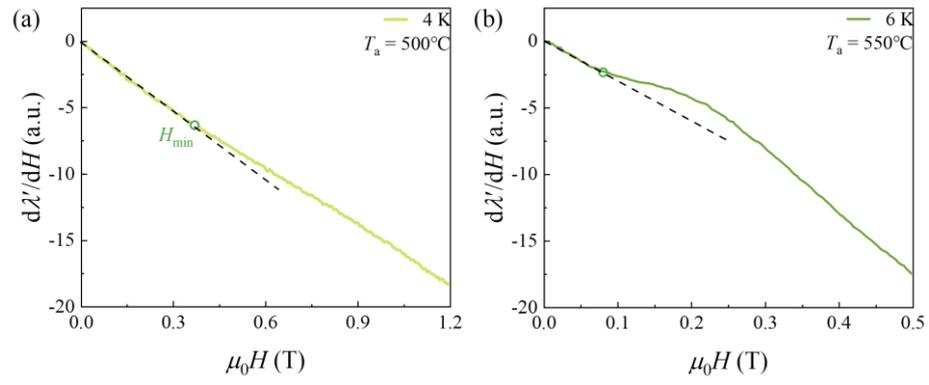

**FIG. S3** Magnetic field dependence of d$\lambda'$/d$H$ at low fields: (a) $T_a$ = 500 °C sample at 4 K; (b) $T_a$ = 550 °C sample at 6 K.